# Large Language Models in Architecture Studio: A Framework for Learning Outcomes

Juan David Salazar Rodriguez[1], Sam Conrad Joyce[1], Nachamma Sockalingam[2], Khoo Eng Tat[3], Julfendi[4]

[1] META Design lab, Architecture and Sustainable Design Pillar, Singapore University of Technology and Design, Singapore, Singapore
[2] Office of Strategic Planning, Singapore University of Technology and Design, Singapore, Singapore
[3] College of Design and Engineering, National University of Singapore, Singapore, Singapore
[4] Engineering Product Development Pillar, Singapore University of Technology and Design, Singapore, Singapore
1008372@mymail.sutd.edu.sg, sam_joyce@sutd.edu.sg, nachamma@sutd.edu.sg, etkhoo@nus.edu.sg, julfendi_julfendi@mymail.sutd.edu.sg

**Abstract.** The study explores the role of large language models (LLMs) in the context of the architectural design studio, understood as the pedagogical core of architectural education. Traditionally, the studio has functioned as an experiential learning space where students tackle design problems through reflective practice, peer critique, and faculty guidance (Schön, 1987; Salama, 2016). However, the integration of artificial intelligence (AI) in this environment has been largely focused on form generation, automation, and representation-al efficiency neglecting its potential as a pedagogical tool to strengthen student autonomy, collaboration, and self-reflection. The objectives of this research were: (1) to identify pedagogical challenges in self-directed, peer-to-peer, and teacher-guided learning processes in architecture studies; (2) to propose AI interventions, particularly through LLM, that contribute to overcoming these challenges; and (3) to align these interventions with measurable learning outcomes using Bloom's taxonomy. The findings show that the main challenges include managing student autonomy, tensions in peer feedback, and the difficulty of balancing the transmission of technical knowledge with the stimulation of creativity in teaching. In response to this, LLMs are emerging as complementary agents capable of generating personalized feedback, organizing collaborative interactions, and offering adaptive cognitive scaffolding. Furthermore, their implementation can be linked to the cognitive levels of Bloom's taxonomy: facilitating the recall and understanding of architectural concepts, supporting application and analysis through interactive case studies, and encouraging synthesis and evaluation through hypo-thetical design scenarios. In conclusion, LLMs constitute an innovative framework for expanding the pedagogical functions of the design studio, repositioning AI as a co-teacher rather than a co-designer. This perspective opens up the possibility of a more autonomous, reflective, and collaborative architectural education, where learning outcomes are not limited to formal production but also encompass the comprehensive training of the designer.

**Keywords:** Large Language models (LLMs), artificial intelligence (AI), pedagogical challenges, architectural design studio.

## 1 Introduction

Architectural design studio has long been the core component of architectural education. The studio has remained an environment for inspiration, critical analysis, and iterative design development throughout recent history, from the master-apprentice traditions of the École des Beaux-Arts to the interdisciplinary and experimental practices of the Bauhaus (Anteet & Binabid, 2025). The studio serves as more than just a place to work; it is a pedagogical setting where students learn by doing, navigating challenging design problems through collaborative work, critique, and reflective practice (Schön, 1987; Salama, 2016).

According to recent research, the studio is a practicum, a semi-professional setting that simulates real-world architectural problems through independent investigation and iterative improvement (Anteet & Binabid, 2025). But the relationships in this learning environment are just as important as the tasks themselves. According to Lodson and Ogbeba (2020), teacher-student dynamics have a significant impact on students' creative performance. While more collaborative or sympathetic profiles encourage a sense of autonomy and confidence, authoritarian or overly assertive tutors may suppress student expression. Peer learning, which frequently takes place in casual, unsupervised settings but is nonetheless rich in knowledge sharing and identity development, is also very important in the studio.

Simultaneously, artificial intelligence (AI) has become a disruptive force in education, offering information-driven decisions, real-time feedback, and personalization. AI has been particularly welcomed in the architectural field due to its potential to enhance the design process. Students can now quickly iterate and visualize ideas thanks to the integration of generative tools like Midjourney, DALL·E, and Stable Diffusion into early ideation phases (Agkathidis et al., 2024; Jin et al., 2024). AI adoption is in line with the larger movement of technology-enhanced design, which aims to free students from technical limitations and enable more creative work.

However, despite the interest surrounding AI in design, its application in education, especially in studio-based pedagogy, is still in its early stages. Recent research indicates that rather than being agents of pedagogical change, AI applications in architecture are primarily presented as tools for form generation, automation, or representational efficiency (Ozorhon et al., 2025; Zahra et al., 2025). In other words, AI has been positioned more as a co-designer than a co-teacher. Therefore, AI-enhanced support has not yet fully benefited critical domains like self-assessment, peer critique, cognitive reflection, or metacognitive development.

These imbalances raises the essential question: is it possible to reposition AI to support both the creation of design and the growth of designers? Constructivism, heutagogy, and student-centered learning theories all support learning environments in which students are independent, introspective, and cooperative (Anteet & Binabid, 2025). AI could serve as a feedback engine that sup-ports critical thinking, a moderator of critique sessions, or a guide for self-reflection if it is developed with pedagogical sensitivity. These opportunities pave the way for a new era of architecture education where AI augments not only output, but outcomes.

By mapping AI-supported learning activities in architectural design studio to Bloom's Taxonomy levels, this paper aims to investigate how AI can support learning outcomes in architecture studio education, beyond design generation.

Chapter 5 organises the actions enabled by LLMs according to the three types of learning. It includes comparative tables showing how AI can support skill development, linking activities, challenges, interventions, and expected learning outcomes.

The paper specifically aims to,

- Identify the pedagogical challenges students face in self-, peer-, and professor-guided learning in architecture studios
- Propose AI interventions that address these challenges
- Align those interventions with measurable learning outcomes using Bloom's Taxonomy

The objective of this paper is to identify the pedagogical challenges students face in self-, peer-, and professor-guided learning in architecture studios. Then propose AI interventions that address these challenges, align those interventions with measurable learning outcomes using Bloom's Taxonomy.

## 2 Methodology

### 2.1 Research design

This conceptual study adopts a qualitative approach to constructing analytical frameworks, with the aim of (1) synthesizing the specialized literature, (2) identifying the main challenges as-sociated with learning in architecture studies, (3) relating the seven basic activities proposed by Kuhn (2001) to the different modes of learning, and (4) establish the correspondence between these challenges and the pedagogical tools based on artificial intelligence (AI) that could contribute to addressing them.
As described in the following phases:

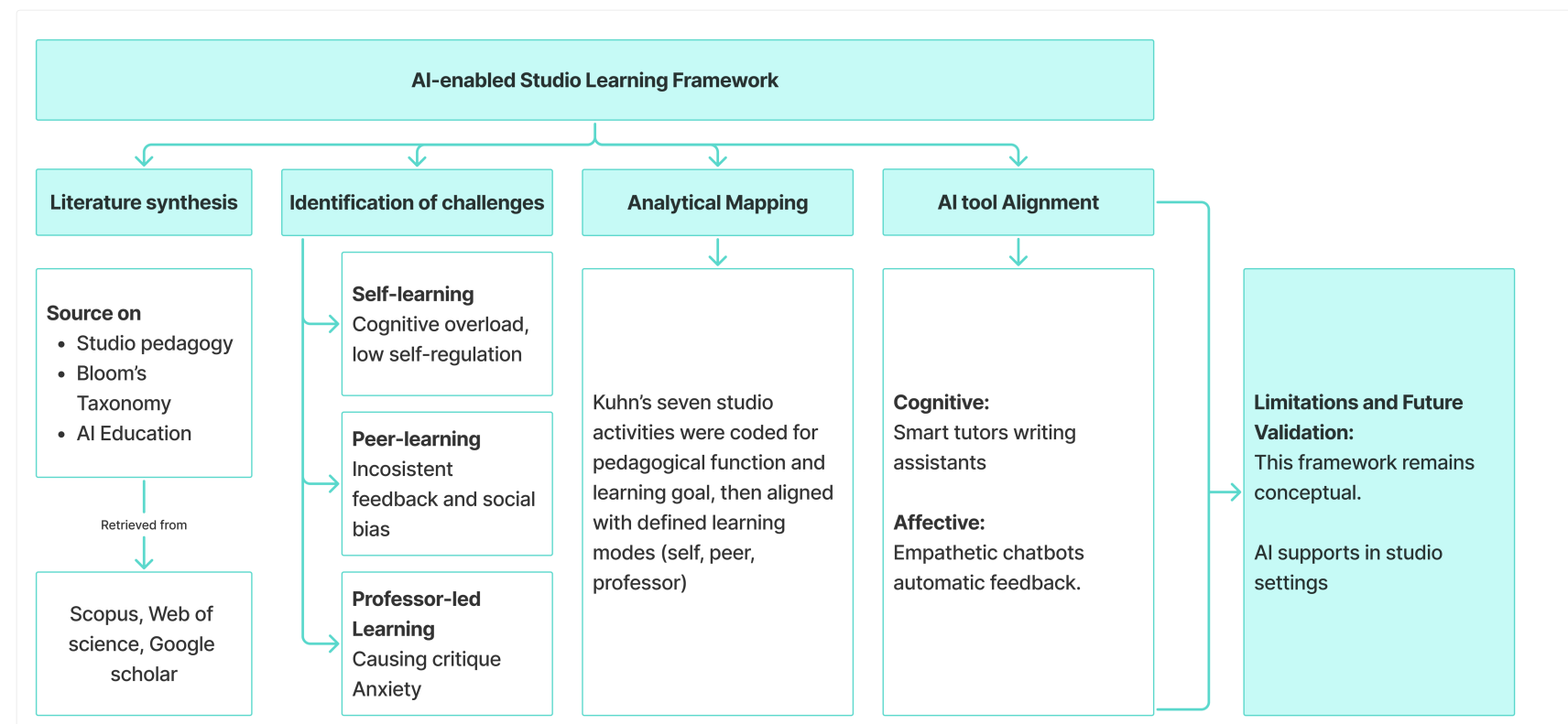

Figure 1. Phased Methodology

Literature Synthesis Phase. A systematic and critical review was conducted of classic and contemporary sources related to study pedagogy (Kuhn, 2001; Schön, 1983), Bloom's taxonomy (Bloom, 2010), and the application of AI in education. To this end, academic databases such as Scopus, Web of Science, and Google Scholar were consulted, using search terms such as "learning in architecture studies," "pedagogical challenges," and "AI in education."

**Identification of learning Challenges Phase.** Both empirical studies and theoretical contributions were analyzed in order to characterize the main pedagogical challenges, grouped into three modes of learning: (a) self-learning (e.g., cognitive overload, low self-regulation), (b) peer learning (e.g., inconsistent feedback, social biases), and (c) teacher-guided learning (e.g., anxiety about criticism, limited guidance). These challenges were classified according to Bloom's cognitive domain levels, which provided the basis for the subsequent selection of pedagogical tools.

**Analytical Mapping Phase.** The seven study activities defined by Kuhn (2001) were coded according to their pedagogical purpose and the learning objectives they address. Subsequently, their correspondence with the three previously defined modes of learning (self-directed, peer-to-peer, and teacher-led) was established. This process was systematized using a coding matrix that allowed for iterative and consistent mapping

**Alignment with AI tools Phase.** Based on a critical analysis of AI-mediated educational systems, tools with evidence of cognitive and/or affective support were identified. The functionalities of these tools were contrasted with the previously mapped challenges and activities, with the aim of generating hypotheses about possible pedagogical interventions to be evaluated in future studies.

It should be noted that the proposed framework is eminently conceptual in nature. Therefore, it is recommended that subsequent research test the identified AI tools in real study contexts, using empirical validation designs based on pre- and post-intervention measurements (e.g., metacognitive awareness inventories, critique rubrics, and peer assessment surveys). These processes will allow for the refinement and validation of the model proposed here

## 3 Gaps and challenges in architecture studio learning

According to Kuhn (2001), the architectural design studio is defined by seven fundamental activities that together influence students' learning and growth as designers. These seven fundamentals activities include open-ended projects that takes a semester long, rapid iterations of design solutions, frequent formal and informal critique, consideration of heterogeneous issues, study of the earlier design examples(precedents) and oscillation between big-picture and detailed thinking, Faculty-guided constraint setting, and the use of diverse design media.

The first activity is the open-ended projects that require students to complete semester-long assignments (also known as “programs”). The students will be given the context of the projects such as the type of building, site of the building, client needs, and the limitations. Due to the unrestricted nature of the projects, students have the flexibility in how to approach and resolve the problem.

After the context is established in the first activity, the students move on to the second activity, which is to perform rapid iterations of design solutions. In this task, students are motivated by the need to present a suggested solution early in the semester. They must swiftly come up with a number of possible solutions (e.g. by sketching). Because of they are expected to be fast, students must think quickly, get early feedback and "size up" (does this mean evaluate?) situations using both explicit and tacit knowledge.

Once the students have drafted their initial solutions, they advanced to the third activity, where they ask for critiques, both formally and informally. This practice creates a "culture of critique" in which students receive feedback from peers, faculty, and outside experts in a variety of ways which include group reviews, final juries, desk crits (discussions at the student's desk), and pin ups (presenting drawings on the wall). While asking for critique is identified as the third activity, critique does not follow a strict order. Instead, it is interwoven across the other activities in Kuhn’s framework and may occur at multiple stages of the design process.

After receiving the initial critiques, students move on to the fourth activity where they consider the heterogeneous issues that emerge from their proposed solutions. Due to their design decisions are interconnected to the brief of their project, initial solutions and received critiques, the discourse of their design involves a variety of concerns (such as social impact, technical viability, structural integrity, aesthetics, and function) frequently in a single discussion.

In analyzing the design issues, students also engage in the fifth activity, which involves studying of previous designs (precedents) and oscillating between thinking holistically and thinking about the detail. Instead of copying the precedents, students examine them to get ideas and encourage rethinking of their problem and solutions. This process requires students to think abstractly about the problem as a whole while also focusing on specific design elements.

Following the study of precedents and oscillations, the sixth activity focuses on how the faculty guides students by setting appropriate constraints. The constraint, such as time, re-sources, and focus areas, help students manage the intricacy of the open-ended problem and arrive at a solution.

After the students are clear on their briefs, constraints, ideas and insight from the critiques, they learn to use of diverse design media to develop their ideas. Proper use of a range of design medias (e.g. sketches, models, software prototypes, site visits, posters, etc) can enhance the insight and design quality. This is because the affordances and constraints of various media allow for various forms of investigation and understanding. Furthermore, the interaction with the media, whether by oneself or with peers, generates unexpected insights (also known as "backtalk") during the "reflective conversation" that is the design process.

Taken together, these seven activities are positioned into Double Diamond framework, modes of learning and Bloom's Taxonomy to respectively illustrate its flow, the people involved in the activities and the cognitive challenges.

The Double diamond is a design process model that describes the divergent and the convergent stages of design process (Tschimmel, K, 2012). It consists of four phases: Discover, Define, Develop, and Deliver. Its iterative nature provides clear visual structure that organizes Kuhn's seven studio activities into a cycle of divergent and convergent. The professor-guided constraints and open-ended briefs align with the **Discover phase**, which is the initial divergent process where the designers usually look for new information and new opportunities. The consideration of heterogeneous issues, precedents, and oscillation between big-picture and detail belong to **Define phase**, which is a convergent stage where students refine and narrow ideas to identify a design opportunity. Activities such as rapid iterations and the use of diverse media fall into **Develop phase**, which is another divergent process, but this time focuses on iterating different ideas and solutions. Both formal and informal critiques occur throughout the process, but the formal critique from the student's final presentations to the final juries falls into the **Deliver phase**.

Beyond their sequence in the design process, Kuhn's activities can also be situated within modes of learning which consist of self-learning, peer-based, or professor-led learning. First, the activities that fit into the self-learning mode are the activities that can be done even without the help of peers or professors. These activities include framing the problem in open-ended projects, performing rapid iteration on their ideas, exploring different medias, examining precedents, and oscillating between big picture and details. Second, the activities for peer-learning are the activities that require interaction between students and their peer. These activities include informal peer critiques, discussion of heterogenous issues and collaborating with peers through varied media. Lastly, the activities for professor-led learning are the activities that require the guidance of the professor. These activities include formal critiques in form of pin-up and jury panel, professor-imposed constraints and structured critique formats.

Once the activities are situated within different modes of learning, they can further be mapped onto Bloom's taxonomy. This framework maps out the cognitive challenges in six levels, from remember, understand, analyze, apply, evaluate to create. In the section that follows, these challenges are discussed within self-learning, peer-learning and professor-led learning.

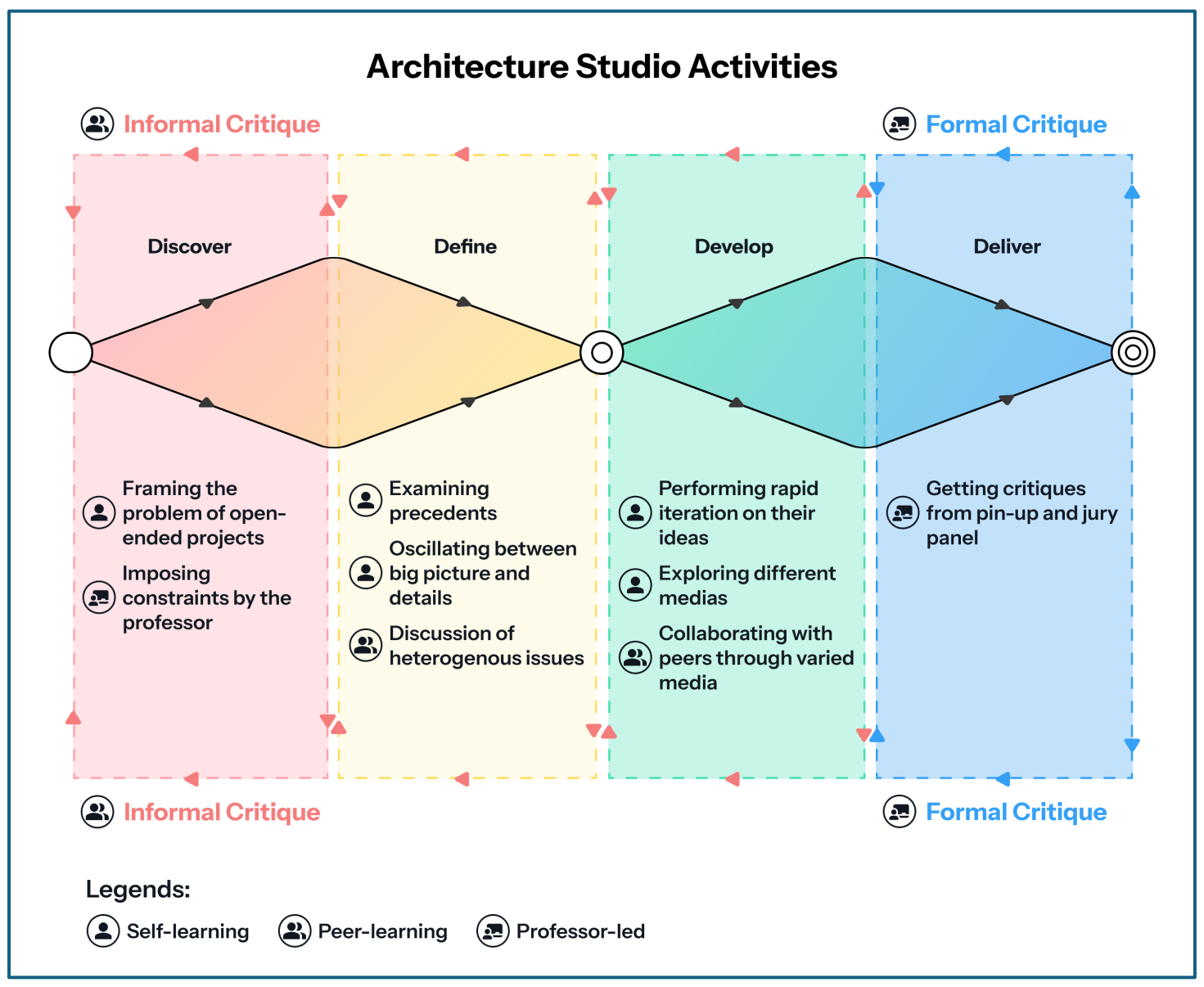

Figure 2. Architecture Studios Activities in the Double Diamond Framework.

### 3.1 Cognitive Challenges in Architecture Studio Activities

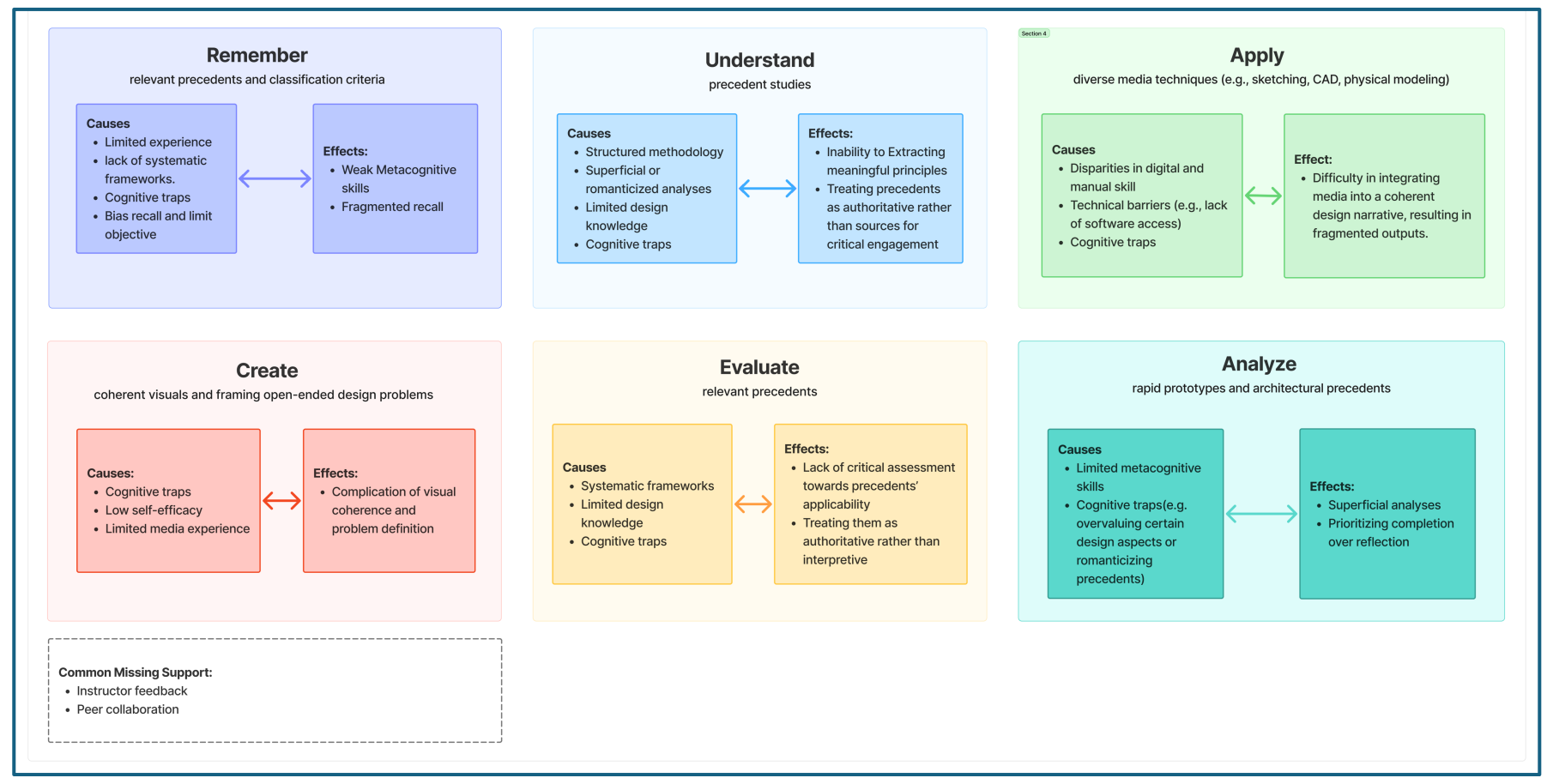

Figure 3. Cognitive Challenge Domain of Self-learning

The cognitive challenges of architectural studio activities for self-learning happens in their study of precedents, oscillating between big picture and details, performing rapid iteration with different medias and framing the problem of their project. When studying the precedents, students struggle to remember, analyze and relevant precedents.

Applying diverse media techniques (e.g., sketching, CAD, physical modeling) is challenging due to disparities in digital and manual skills, technical barriers (e.g., lack of software access), and cognitive traps like over-reliance on familiar techniques (Asadpour, 2021; Avsec & Jagiełło-Kowalczyk, 2021). Without feedback, students struggle to integrate media into a coherent design narrative, resulting in fragmented outputs.

Analyzing rapid prototypes and architectural precedents is difficult due to limited metacognitive skills and cognitive traps, such as overvaluing certain design aspects or romanticizing precedents (Avsec & Jagiełło-Kowalczyk, 2021; Yaseen et al., 2022; Alassaf, 2025). The absence of structured feedback leads to superficial analyses, while the fast-paced nature of prototyping prioritizes completion over reflection (Liu et al., 2020; Yang & Kim, 2024).

Evaluating relevant precedents is challenging due to the lack of systematic frameworks, limited design knowledge, and cognitive traps like overvaluing visually appealing precedents (Yaseen et al., 2022; Alassaf, 2025; Avsec & Jagiełło-Kowalczyk, 2021). Students often fail to critically assess precedents' applicability, treating them as authoritative rather than interpretive (White, 2024).

Creating coherent visuals and framing open-ended design problems are hindered by cognitive traps (e.g., rigid thinking, focusing on familiar techniques), low self-efficacy, and limited media experience (Avsec & Jagiełło-Kowalczyk, 2021; Sha et al., 2024; Asadpour, 2021). The absence of feedback and technical barriers (e.g., lack of software access) further complicates visual coherence and problem definition (Asadpour, 2021).

### 3.2 Peer Learning Challenges in Architecture Studio Activities

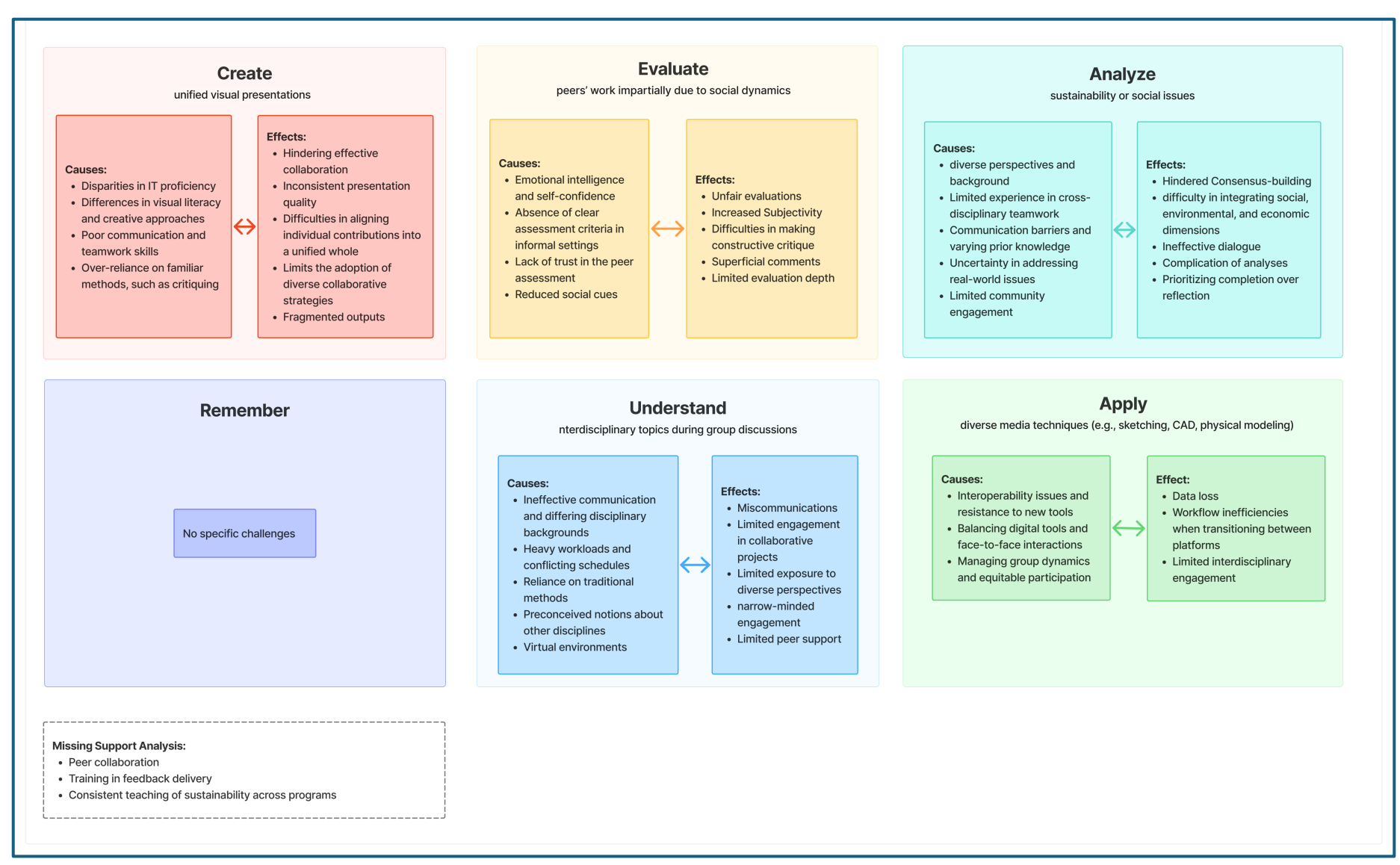

Figure 4. Cognitive Challenge Domain of Peer-learning

Architecture students face significant barriers when creating unified visual presentations during peer learning due to varied media skills. Disparities in IT proficiency hinder effective collaboration, especially in virtual environments where peer-to-peer support is limited (Wang, 2023). Differences in visual literacy and creative approaches often lead to inconsistent presentation quality, as students may prioritize technical skills over cohesive composition (Stanimirović et al., 2023). Poor communication and teamwork skills further complicate aligning individual contributions into a unified whole (Mohammad Yusoff et al., 2022). Over-reliance on familiar methods, such as critiquing, limits the adoption of diverse collaborative strategies (Lotfabadi & Iranmanesh, 2023). The lack of structured peer support mechanisms exacerbates fragmented outputs, necessitating intentional development of both technical and soft skills (Wang, 2023; Tafahomi, 2021).

During informal desk critiques, architecture students struggle to evaluate peers' work impartially due to social dynamics, such as friendships and group hierarchies, which introduce bias (Sinaga et al., 2024). Emotional intelligence and self-confidence affect the fairness of evaluations, as students may over- or underestimate their own or others' work (Sinaga et al., 2024). The absence of clear assessment criteria in informal settings increases subjectivity, while a lack of trust in the peer assessment process discourages honest feedback (Sinaga et al., 2024). Virtual environments further complicate this by reducing social cues, making constructive critique hard-er (Wang, 2023). Without training in feedback delivery, students often provide superficial comments, limiting evaluation depth (Sinaga et al., 2024; Inam, 2025).

When analysing sustainability or social issues, students face challenges due to diverse perspectives and backgrounds, which hinder consensus-building (Keogh, 2021; Avsec et al., 2022). Limited experience in cross-disciplinary teamwork makes it difficult to integrate social, environ-mental, and economic dimensions (Keogh, 2021; Calikusu et al., 2022). Communication barriers and varying prior knowledge impede effective dialogue (Núñez-Andrés et al., 2021; Campbell et al., 2024). Uncertainty in addressing real-world issues, coupled with limited community engagement, further complicates analysis (De Wit-Paul, 2024; Avsec et al., 2022). Inconsistent teaching of sustainability across programs adds to these challenges (Boarin et al., 2020; Calikusu et al., 2022).

Collaborating on shared projects using mixed digital and physical media tools presents technical challenges, such as data loss and workflow inefficiencies when transitioning between platforms (El-Khouly & Abdelhalim, 2024). Interoperability issues and resistance to new tools hinder teamwork (Jin et al., 2018). Balancing digital tools' benefits with face-to-face interaction is difficult, as is managing group dynamics and equitable participation (Leicht et al., 2009; Liu et al., 2024; Emam et al., 2019). Traditional teaching methods further impede interdisciplinary engagement (Indraprastha, 2023).

Understanding interdisciplinary topics during group discussions is hindered by ineffective communication and differing disciplinary backgrounds, leading to misunderstandings (Keenahan & McCrum, 2020). Heavy workloads and conflicting schedules limit engagement in collaborative projects (Ali, 2019). Virtual environments reduce access to peer support, while reliance on traditional methods like critiquing limits exposure to diverse perspectives (Wang, 2023; Lotfabadi & Iranmanesh, 2023). Preconceived notions about other disciplines further impede open-minded engagement (Ali, 2019).

This study does not provide specific challenges for the "Remember" level, suggesting that foundational knowledge recall is less problematic in peer learning compared to higher-order cognitive skills. However, this could imply that students may struggle to retain interdisciplinary or collaborative knowledge without structured reinforcement.

### 3.3 Professor-Led Learning Challenges in Architecture Studio Activities

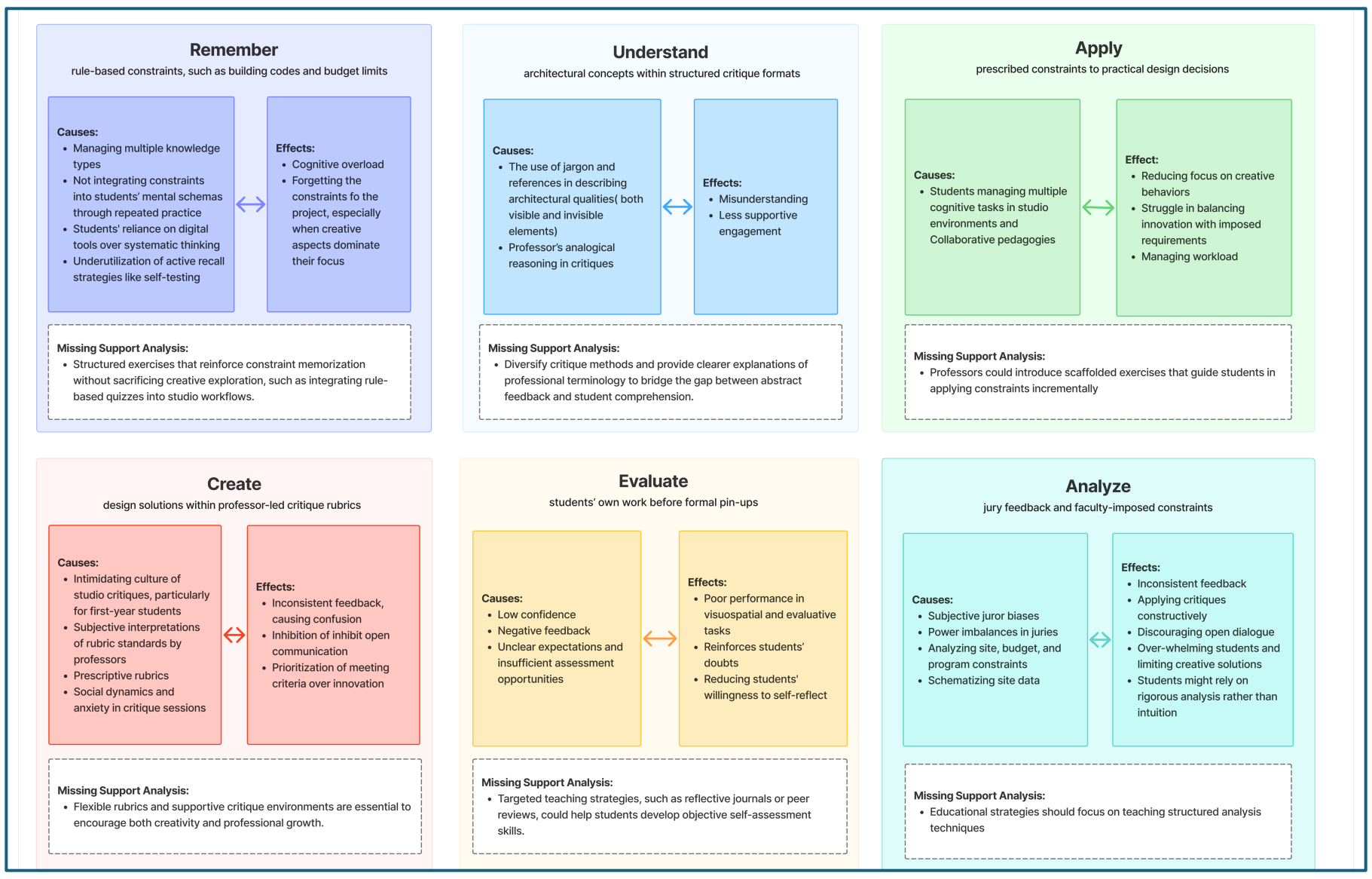

Figure 5. Domain of Cognitive Challenge for Professor-led Learning

At the Remembering level, students struggle to recall rule-based constraints, such as building codes and budget limits, during professor-led assignments. The complexity of design problems requires students to manage multiple knowledge types, often leading to cognitive overload. According to Chan (1990), if constraints are not integrated into students' mental schemas through repeated practice, they are easily forgotten, especially when creative aspects dominate their focus. Yazıcı (2019) highlights that students' reliance on digital tools over systematic thinking exacerbates this issue, as does the underutilization of active recall strategies like self-testing (Xu et al., 2024). This challenge suggests a need for structured exercises that reinforce constraint memorization without sacrificing creative exploration, such as integrating rule-based quizzes into studio workflows.

Understanding architectural concepts within structured critique formats is challenging due to the complexity of professional language and visual practices. Lymer (2009) notes that the use of jargon and references to implicit standards can confuse students, particularly when architectural qualities are discussed as both visible and invisible elements. Murphy et al. (2012) add that analogical reasoning in critiques often fails to connect with students still developing their conceptual frameworks. Additionally, student preferences for critique formats, such as desk critiques versus pin-up critiques, influence engagement, with some formats feeling less supportive (Gul & Afa-can, 2018). This suggests that professors should diversify critique methods and provide clearer explanations of professional terminology to bridge the gap between abstract feedback and student comprehension.

Applying prescribed constraints to practical design decisions is difficult, as constraints shift students' cognitive focus from problem framing to simultaneous problem-solving. Ashrafganjouei & Gero (2020) explain that visual constraints complicate workflows by forcing students to manage multiple cognitive tasks, reducing focus on creative behaviors. Kuhn (2001) notes that while studio environments encourage creative constraint use, students struggle to balance innovation with imposed requirements. Collaborative pedagogies can help, but managing workload remains a challenge (Xhambazi & Aliu, 2024). To address this, professors could introduce scaffolded exercises that guide students in applying constraints incrementally, fostering both practical skills and creative confidence.

Students face significant hurdles in analyzing and integrating jury feedback and faculty-imposed constraints. Al-Qemaqchi (2024) points out that subjective juror biases lead to inconsistent feedback, making it hard for students to apply critiques constructively. Yorgancıoğlu et al. (2021) highlight power imbalances in juries, which discourage open dialogue. Similarly, analyzing site, budget, and program constraints is challenging due to their complexity, often overwhelming students and limiting creative solutions (Pauwels et al., 2014). Park et al. (2023) note difficulties in schematizing site data, while Osmólska & Lewis (2023) suggest students rely on intuition rather than rigorous analysis. Educational strategies should focus on teaching structured analysis techniques, such as constraint mapping, to help students balance compliance and creativity.

Evaluating their own work before formal pin-ups is undermined by limited metacognitive awareness and self-efficacy. Acar (2021) explains that low confidence leads to poor performance in visuospatial and evaluative tasks, while Acar & Acar (2020) highlight the impact of individual differences, such as visuospatial abilities. Negative feedback in studio settings can reinforce doubts, reducing students' willingness to self-reflect (Acar, 2021). Mclaughlan & Chatterjee (2020) note that unclear expectations and insufficient assessment opportunities further hinder self-evaluation. Targeted teaching strategies, such as reflective journals or peer reviews, could help students develop objective self-assessment skills.

Creating design solutions within professor-led critique rubrics is challenging due to the intimidating culture of studio critiques, particularly for first-year students (Hamza et al., 2023). Prescriptive rubrics can constrain creativity, as students prioritize meeting criteria over innovation (El-Latif et al., 2020). Subjective interpretations of rubric standards by professors lead to inconsistent feedback, causing confusion (Megahed, 2017). Social dynamics and anxiety in critique sessions further inhibit open communication (Yorgancıoğlu & Tunalı, 2020). Flexible rubrics and supportive critique environments are essential to encourage both creativity and professional growth.

# 4. LLMs Actions and Learning Outcomes

With the challenges across multiple levels, LLMs can be used to reshape the architecture studio education by supporting self-learning, peer-learning and professor-led activities.

## 4.1 Self-Learning Actions Enabled by LLMs in Architecture Studios

LLMs strengthen systematic precedent recall by converting textual and visual data into structured prompts that aid long-term memory and reduce bias toward visually striking examples. This cognitive reinforcement allows students to interpret precedents through conceptual rather than aesthetic similarity (Kaithe et al., 2025; Matzakos & Moundridou, 2025). Parallelly, resilience training in ambiguity uses adaptive prompting to build emotional intelligence and tolerance for uncertain design problems, an ability linked to deeper reflection and iterative reasoning (Wang et al., 2023; Liu et al., 2023).

LLMs promote critical-analysis scaffolding by asking targeted "why/how" questions that emulate expert feedback and enhance confidence in reasoning through precedent interpretation (Kumar et al., 2024; Ali et al., 2025). They also serve as design-validation systems, prompting students to test their assumptions against evidence, contrast alternatives, and articulate decision logic through rubric-based reflection (Louatouate & Zeriouh, 2025; Matzakos & Moundridou, 2025).

Models enable personalized learning for design software, delivering just-in-time tutorials and multimodal explanations that connect tool proficiency with conceptual goals. By filling digital-skill gaps, these systems ensure equitable learning progress and facilitate translation between sketches, parametric models, and simulations (Li et al., 2024; Yue et al., 2024).

LLMs support visual-storytelling development and structured-evaluation frameworks. Students can generate coherent narrative scripts for their visual outputs, ensuring alignment between design intent and representation (Gao et al., 2024; Miao et al., 2024). Concurrently, automated evaluation frameworks reduce subjective bias by standardizing feedback language and highlighting recurring conceptual patterns (Wang et al., 2025).

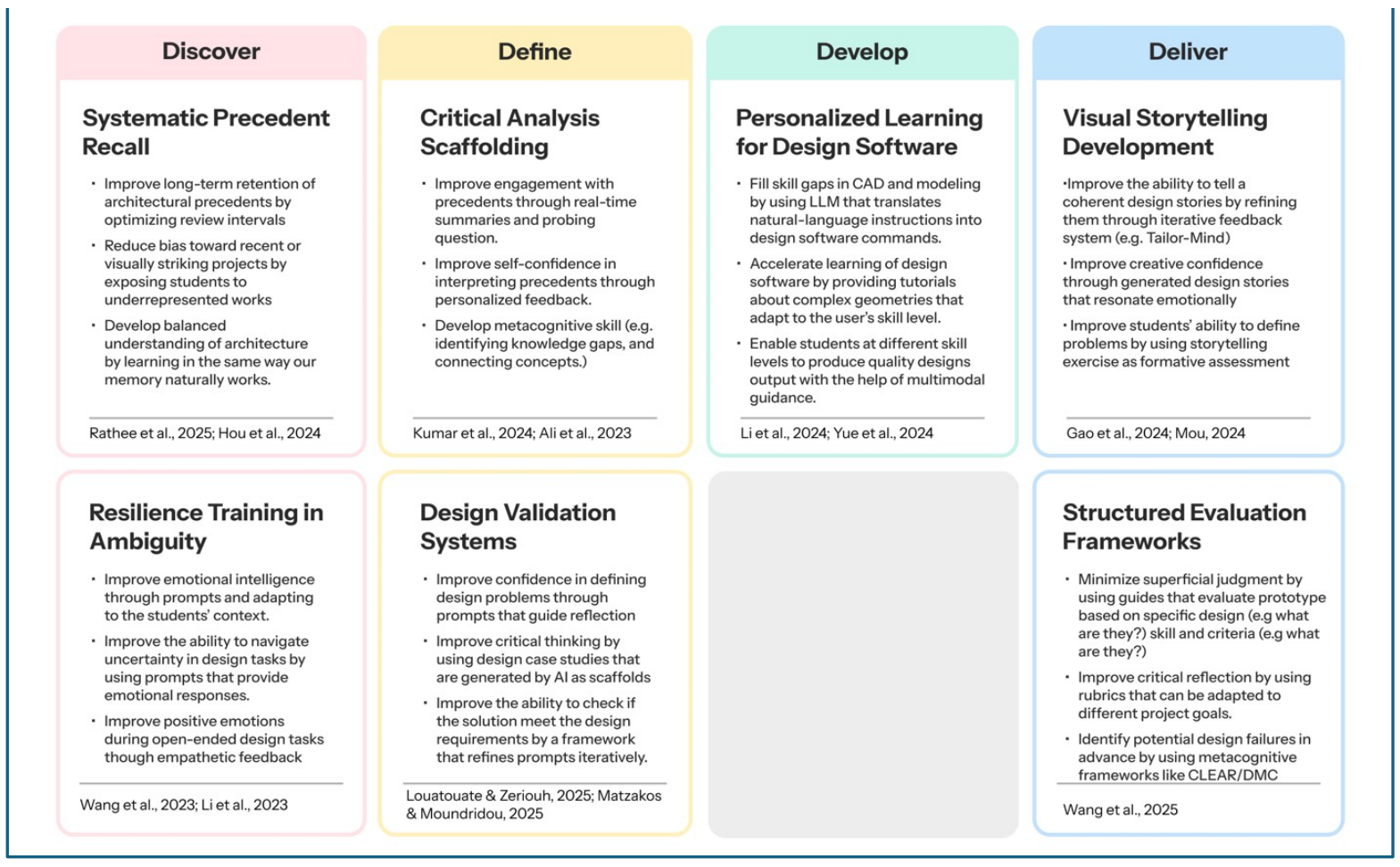

Figure 6. Applications of LLMs for Self-learning in architecture studio

## 4.2 Peer-Learning Actions Enabled by LLMs in Architecture Studios

LLMs support virtual peer simulation by orchestrating debate scenarios where students defend or critique design decisions in real time. These simulated exchanges improve argumentation, metacognitive reflection, and reduce anxiety before live juries (Liu et al., 2024; Kjærsdóttir & Tzortzopoulos, 2024). Through adaptive tutoring, models provide just-in-time prompts that guide peers toward constructive dialogue and evidence-based critique, strengthening reasoning and teamwork when preparing group presentations (Groscurth et al., 2015; Fu et al., 2024).

A recurring issue in studio collaboration is bias rooted in friendship networks, hierarchy, or differing creative preferences. LLMs can mitigate this through bias-mitigated feedback, standardizing critique language and detecting emotional polarity to ensure balanced exchanges (Fan et al., 2024; Mohanty, 2025). Moreover, value-conflict navigation routines train students to debate ethically and articulate trade-offs between functionality, aesthetics, and sustainability, helping them reconcile opposing design philosophies (Aggrawal & Magana, 2024; Zhang et al., 2025).

In the critique process, feedback-integration systems allow peers to merge multiple perspectives through semantic clustering and topic mapping, helping identify overlaps and contradictions across discussions (Cebrian & Doyle, 2024; Zhou et al., 2025). LLMs also provide structured-feedback scaffolding, prompting students to justify opinions with analytical reasoning rather than preference. By translating qualitative comments into categorized rubrics, models foster higher-order evaluation and argument coherence (Guo, 2024; Uyarel et al., 2024).

Finally, rubric-driven assessment frameworks use LLMs to calibrate peer evaluations against instructor rubrics such as CLEAR or PMC, promoting consistency and fairness. Students learn to align their feedback with shared quality indicators while reflecting on their own evaluative thinking (Fan et al., 2025; Kimet et al., 2023). This transforms assessment into a bidirectional learning process that benefits both reviewer and recipient.

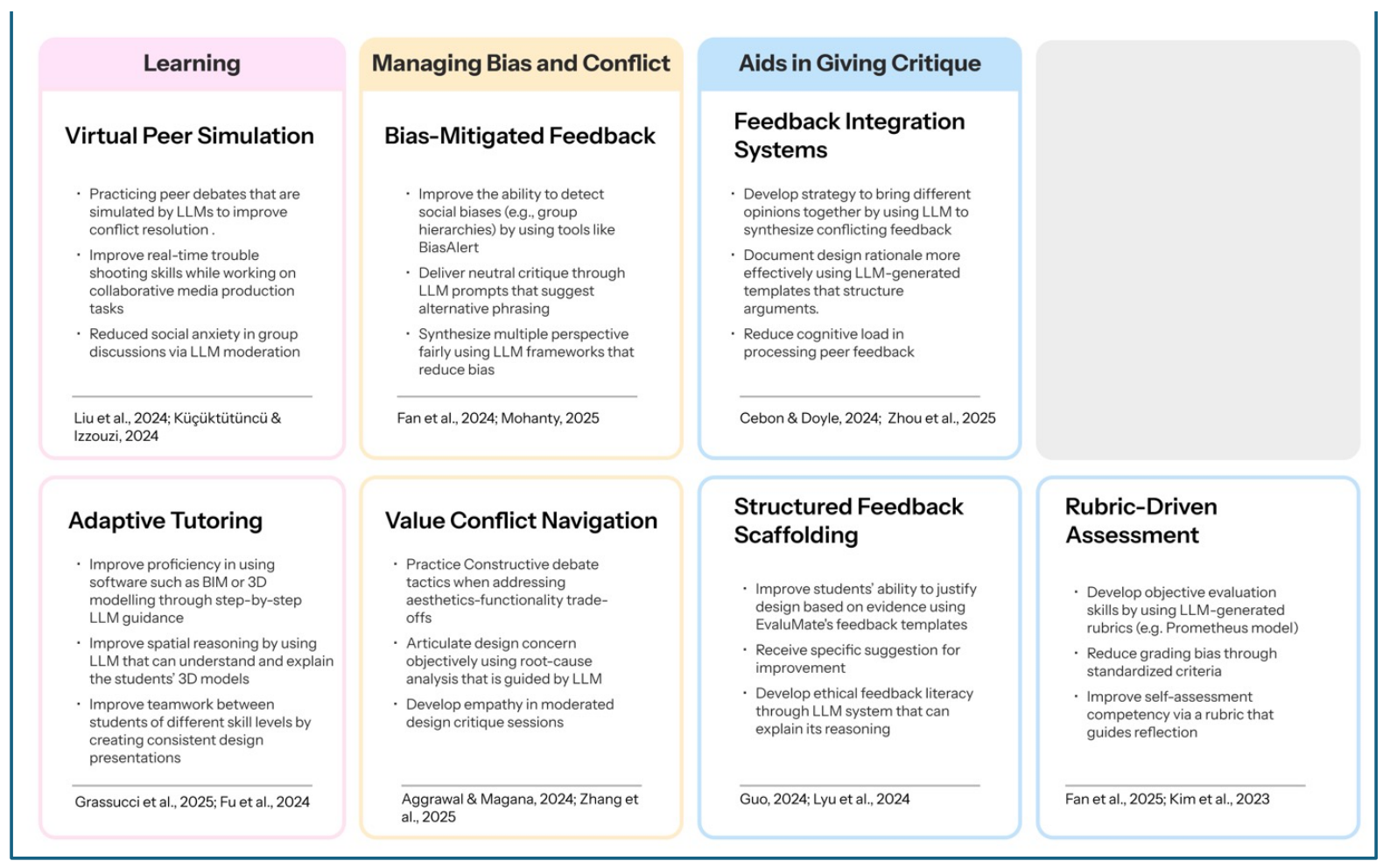

Figure 7. Applications of LLMs for Peer-learning in architecture studio

### 4.3 Professor-led Actions Enabled by LLMs in Architecture Studios

Within professor-led studio settings, Large Language Models (LLMs) serve as pedagogical co-facilitators that augment the instructor's feedback capacity, promote clarity in communication, and enhance students' reflective understanding. As depicted in Figure 8, these models strengthen the entire feedback cycle—from critique delivery to evaluative reasoning—across four pedagogical domains.

Aids in Giving Critique

LLMs can support ethical reasoning prompts that help professors and students document design logic, justify decisions, and align feedback with ethical and contextual considerations (Zhou et al., 2025; Ollion et al., 2024). Through structured questioning, these systems guide teachers to articulate professional standards and ensure students critically evaluate design consequences. Complementarily, creativity-focused critique modules enable the instructor to generate feedback that identifies originality and divergent thinking, encouraging design exploration rather than conformity (Yang et al., 2025; Kim & Oh, 2025). These affordances make feedback both reflective and formative, rather than merely evaluative.

Architectural feedback often suffers from disciplinary jargon and implicit meaning. LLMs can operate as jargon translation systems, converting technical critique into accessible, student-friendly language while maintaining terminological accuracy (Fan et al., 2024; Muthers, 2025). By pairing analogical reasoning with contextual examples, these systems bridge the gap between professional vocabulary and novice understanding. Likewise, ambiguous feedback interpreters assist both faculty and students in decoding unclear or contradictory remarks, synthesizing comments into actionable insights that promote critical reflection (Aggrawal & Magana, 2024; Zhang et al., 2025).

In the learning and practice phase, LLMs deliver guides for designing within constraints, helping professors illustrate how limitations (budget, codes, climate) can fuel creativity rather than restrict it. By generating adaptive scenarios and problem-based exercises, models scaffold applied reasoning aligned with course objectives (Liu et al., 2024; Kjærsdóttir & Izzoud, 2024). In parallel, metacognitive assistants support individualized tutoring by monitoring students' evolving reasoning patterns and recommending exercises that enhance self-regulation and conceptual depth (Groscurth et al., 2015; Fu et al., 2024).

Finally, jury simulation platforms enable professors to conduct mock critiques where students rehearse argumentation and response strategies using LLM-generated juror personas. These simulations prepare students for formal reviews by developing confidence, adaptive reasoning, and rhetorical precision (Fan et al., 2025; Kim et al., 2023). Such virtual juries also allow instructors to model constructive critique behaviors, reinforcing professional communication and empathy in evaluative contexts.

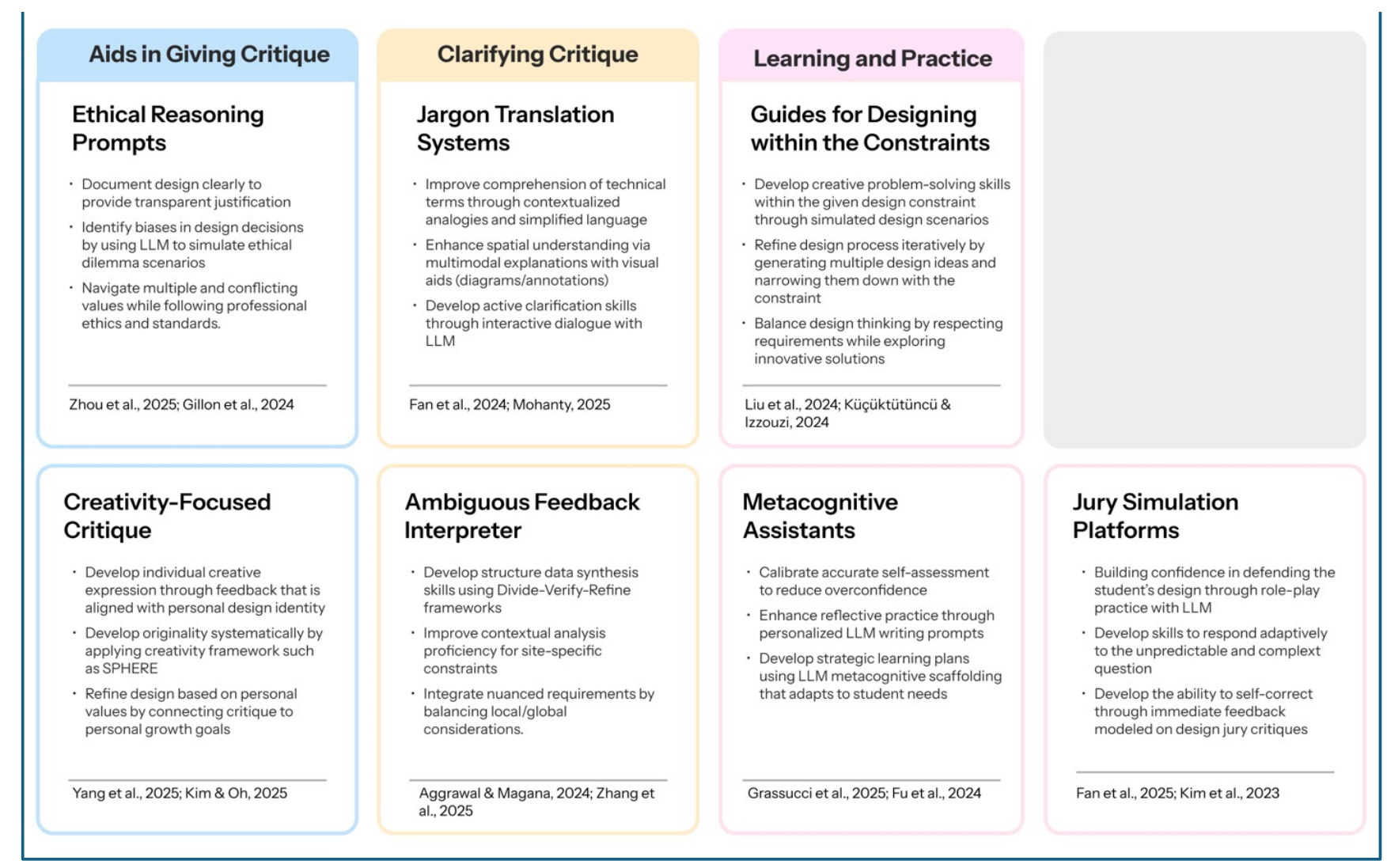

Figure 8. Applications of LLMs for Professor-led learning in architecture studio

## 5 Discussion

The study on the use of large language models (LLMs) in architecture studies opens up a field of pedagogical reflection that allows us to link the potential of artificial intelligence (AI) with learning processes in complex academic contexts. The results obtained allow us to discuss three main areas: pedagogical challenges, the possibilities for AI intervention, and the alignment of these with measurable learning outcomes under Bloom's taxonomy.

First, the pedagogical challenges identified in architecture students are related to the inter-disciplinary and creative nature of their training. Self-directed learning requires high levels of autonomy and critical thinking, which creates difficulties in information management and knowledge self-regulation (Zimmerman, B. 2002). On the other hand, peer learning, although it encourages collaboration and the collective construction of knowledge, faces tensions related to coordination and the quality of feedback (Boud, D., Cohen, R., & Sampson, J. 2014). Finally, in teacher-guided learning, the central challenge lies in the need to balance the transmission of technical knowledge with the promotion of creativity and critical problem solving (Schön, 1983).

In the face of these challenges, LLMs offer significant potential for designing pedagogical interventions. For example, in self-directed learning, AI can act as a cognitive tutor, providing adaptive explanations and architectural simulations that support spatial and conceptual reasoning (Holmes, W., Bialik, M., & Fadel, C. 2019). In peer learning, AI-based tools can serve as mediators, organizing discussions, suggesting relevant academic references, and ensuring that interactions remain aligned with course objectives. In teacher-guided learning, LLMs can contribute to the personalization of feedback, freeing up teacher time to focus on more complex aspects of creative and project-based teaching.

The discussion becomes more robust when these interventions are aligned with Bloom's taxonomy levels. At the lower cognitive level, LLMs facilitate the recall and understanding of architectural concepts through automated summaries and contextual explanations. At intermediate levels, they can promote application and analysis through interactive case studies and the generation of comparisons between design approaches. Finally, at higher levels, LLMs become catalysts for synthesis and evaluation, offering hypothetical design scenarios that allow students to explore innovative solutions and critically argue their decisions.

Kuhn’s Seven Architectural Studio Activities

Kuhn (2001) identifies seven fundamental activities in architecture studios that are crucial to understanding learning processes:

1. Open semester projects.
2. Rapid design iterations.
3. Formal and informal critiques.
4. Consideration of heterogeneous problems.
5. Study of precedents and global-detailed thinking.
6. Teacher-guided constraints.
7. Use of diverse design media.

These activities are associated with three types of learning: self-learning, peer-learning, and teacher-led learning, relating pedagogical challenges and the potential for intervention by LLMs.

The following table summarizes the main challenges and expected learning outcomes for each type of learning:

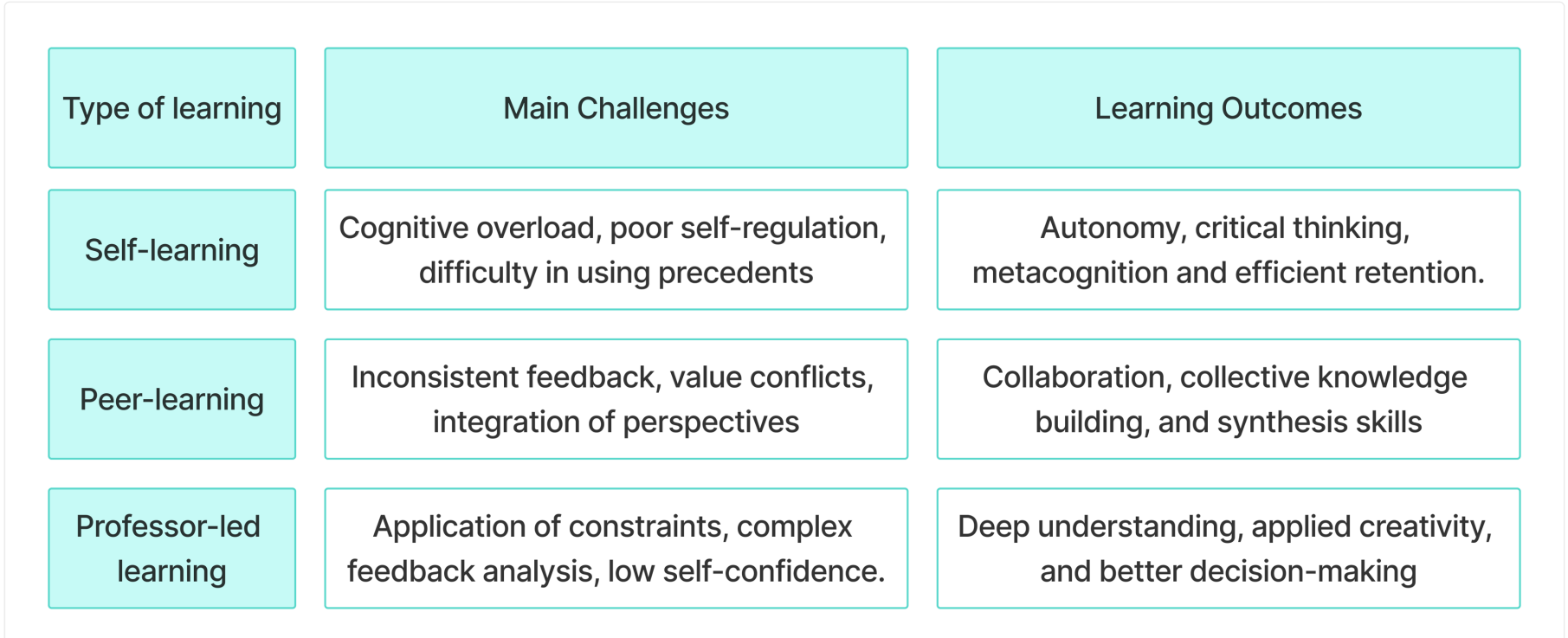

| Type of learning | Main Challenges | Learning Outcomes |
| --- | --- | --- |
| Self-learning | Cognitive overload, poor self-regulation, difficulty in using precedents | Autonomy, critical thinking, metacognition and efficient retention. |
| Peer-learning | Inconsistent feedback, value conflicts, integration of perspectives | Collaboration, collective knowledge building, and synthesis skills |
| Professor-led learning | Application of constraints, complex feedback analysis, low self-confidence. | Deep understanding, applied creativity, and better decision-making |

Figure 9. Main challenges and learning outcomes

## 7 Conclusions

The incorporation of large-scale linguistic models into architectural education does not replace human interaction or design experience, but it does provide a robust framework to support learning processes. Its true value lies in its ability to integrate with existing pedagogical dynamics, promoting autonomy, collaboration, and guided teaching under an evaluation scheme consistent with Bloom's taxonomy.

On the other hand, self-directed, peer-to-peer, and teacher-guided learning in architecture studies faces barriers related to autonomy, the quality of feedback, and the integration of creativity with technical knowledge. These challenges confirm the need for support strategies that promote self-regulation, meaningful collaboration, and reflective teaching.

Similarly, this research shows that LLMs are a promising tool for complementing teaching and learning processes in architecture. Their value lies in their ability to provide personalized feedback, organize collaborative discussions, and facilitate adaptive explanations that strengthen both conceptual understanding and creative exploration.

Finally, the integration of LLMs in the context of architectural education does not aim to re-place the role of the teacher, but rather to strengthen it, freeing up time for creative guidance and reflective support. This synergy between technology and pedagogy promotes a more equitable and personalized learning environment that is aligned with expected learning outcomes.

For this reason, the responsible implementation of LLMs in architectural education requires further research on ethical issues, biases, and technological accessibility. However, their gradual incorporation offers a strategic opportunity to transform teaching environments, expand student autonomy, and enrich the collective construction of knowledge in the discipline.

The incorporation of LLMs into architectural education redefines the dynamics of learning. For self-learning, they facilitate self-regulation through personalised tutoring and the generation of adaptive scenarios. In peer-learning, they serve as neutral mediators that structure feedback and reduce social biases. In teacher-led learning, they complement the teaching role by providing simulations, explanations and assessments tailored to the student's context.

The integration of these technologies creates synergy between the three learning modalities, allowing complex challenges to be addressed and more effective results to be achieved

**Disclosure of interests.** The authors have no competing interests to declare that are relevant to the content of this article.

**Credit author statement.** Juan David Salazar Rodriguez: Conceptualization, Methodology, Investigation, Writing – Original Draft, Visualization. **Sam Conrad Joyce:** Supervision, Writing – Review & Editing. **Nachamma Sockalingam:** Supervision, Writing – Review & Editing. **Eng Tat Khoo:** Supervision, Writing – Review & Editing. **Julfendi:** Writing – Review & Editing.

## References

- Acar, A. (2021). A Research on the relationship between problem-solving appraisal, attention, and visuospatial skills of first-year architecture students. International Journal of Art & Design Education. 18
- Acar, A., & Acar, A. S. (2020). Neuropsychological Assessment of First-Year Architecture Stu-dents' Visuospatial Abilities: Overview. International Journal of Art & Design Educa-tion
- Aggrawal, S., & Magana, A. J. (2024). Teamwork conflict management training and conflict resolution practice via large language models. Future Internet, 16(5), 177.
- Agkathidis, A., Song, Y., & Ioanna, S. (2024). AI-Assisted Design: Utilising artificial intelligence as a generative form-finding tool in architectural design studio teaching.
- Al-Qemaqchi, N. T. (2025). Biases shown by juries during architecture design assessment sessions. European Journal of Engineering Education, 50(4), 724-743.
- Alassaf, N. (2025). Computational Precedent-Based Instruction (CPBI): Integrating Precedents and BIM-Based Parametric Modeling in Architectural Design Studio. Buildings, 15(8), 1287.
- Ali, F., Choy, D., Divaharan, S., Tay, H. Y., & Chen, W. (2023). Supporting self-directed learning and self-assessment using TeacherGAIA, a generative AI chatbot application: Learning approaches and prompt engineering. Learning: Research and Practice, 9(2), 135–147. https://doi.org/10.1080/23735082.2023.2258886
- Ali, A. K. (2019). A case study in developing an interdisciplinary learning experiment between architecture, building construction, and construction engineering and management education. Engineering, Construction and Architectural Management, 26(9), 2040-2059.

- Anteet, Q., & Binabid, J. (2025). Investigating the discourse on pedagogical effectiveness in the architectural design studio. Architectural Engineering and Design Management, 21(1), 17–34. https://doi.org/10.1080/17452007.2024.2402055
- Ashrafganjouei, M., & Gero, J. (2020). Exploring the effect of a visual constraint on students' design cognition. Artificial Intelligence for Engineering Design, Analysis and Manufacturing.
- Asadpour, A. (2021). Student challenges in online architectural design courses in Iran during the COVID-19 pandemic. E-learning and Digital Media, 18(6), 511-529.
- Avsec, S., Jagiełło-Kowalczyk, M., & Żabicka, A. (2022). Enhancing transformative learning and innovation skills using remote learning for sustainable architecture design. Sustainability, 14(7), 3928.
- Boarin, P., Martinez-Molina, A., & Juan-Ferruses, I. (2020). Understanding students' perception of sustainability in architecture education: A comparison among universities in three different continents. Journal of Cleaner Production, 248, 119237.
- Bloom, B. S. (2010). *A taxonomy for learning, teaching, and assessing: A revision of Bloom's taxonomy of educational objectives*. Longman.
- Boud, D., & Cohen, R. (2014). Peer learning in higher education: Learning from and with each other. Routledge.
- Chan, C. S. (1990). Cognitive processes in architectural design problem solving. Design studies, 11(2), 60-80.
- Calikusu, A. N., Cakmakli, A. B., & Gursel Dino, I. (2023). The impact of architectural design studio education on perceptions of sustainability. Archnet-IJAR: International Journal of Architectural Research, 17(2), 375-392.
- De Wit-Paul, A. (2024). Training Architects as Activists: Social Sustainably in the Studio. International Journal on Engineering, Science and Technology, 6(1), 77-88.
- El-Latif, M. A., Al-Hagla, K. S., & Hasan, A. E. (2020). Overview on the criticism process in architecture pedagogy.
- El-Khouly, T., & Abdelhalim, O. (2024). Preserving conceptual design integrity: strategies for enhancing interoperability in architectural digital design workflows. Scientific Reports, 14(1), 30595.
- Emam, M., Taha, D., & ElSayad, Z. (2019). Collaborative pedagogy in architectural design studio: A case study in applying collaborative design. Alexandria Engineering Journal, 58(1), 163-170.
- Fan, Z., Chen, R., Xu, R., & Liu, Z. (2024). Biasalert: A plug-and-play tool for social bias detection in llms. arXiv preprint arXiv:2407.10241.
- Fan, Z., Wang, W., Wu, X., & Zhang, D. (2025). Sedareval: Automated evaluation using self-adaptive rubrics. arXiv preprint arXiv:2501.15595.
- Fu, R., Liu, J., Chen, X., Nie, Y., & Xiong, W. (2024). Scene-llm: Extending language model for 3d visual understanding and reasoning. arXiv preprint arXiv:2403.11401.

- Gao, L., Lu, J., Shao, Z., Lin, Z., Yue, S., Leong, C., ... & Chen, S. (2024). Fine-tuned large language model for visualization system: A study on self-regulated learning in education. IEEE Transactions on Visualization and Computer Graphics, 31(1), 514-524.
- Gunday Gul, C. G., & Afacan, Y. (2018). Analysing the effects of critique techniques on the success of interior architecture students. International Journal of Art & Design Education, 37(3), 469-479.
- Guo, K. (2024). EvaluMate: Using AI to support students' feedback provision in peer assessment for writing. Assessing Writing, 61, 100864.
- Holmes, W., Bialik, M., & Fadel, C. (2019). Artificial intelligence in education promises and implications for teaching and learning. Center for Curriculum Redesign.
- Indraprastha, A. (2023). Fostering Critical Collaborative Thinking through Digital Platform: An Empirical Study on Interdisciplinary Design Project. International Journal of Built Environment and Scientific Research.
- Jin, R., Yang, T., Piroozfar, P., Kang, B. G., Wanatowski, D., Hancock, C. M., & Tang, L. (2018). Project-based pedagogy in interdisciplinary building design adopting BIM. Engineering, Construction and Architectural Management, 25(10), 1376-1397.
- Jin, S., Tu, H., Li, J., Fang, Y., Qu, Z., Xu, F., ... & Lin, Y. (2024). Enhancing architectural education through artificial intelligence: a case study of an AI-assisted architectural programming and design course. Buildings, 14(6), 1613.
- Keogh, S. Embedded and Hopeful: A Curriculum for Change.
- Keenahan, J., & McCrum, D. (2021). Developing interdisciplinary understanding and dialogue between Engineering and Architectural students: design and evaluation of a problem-based learning module. European Journal of Engineering Education, 46(4), 575-603.
- Kim, S., & Oh, D. (2025). Evaluating Creativity: Can LLMs Be Good Evaluators in Creative Writing Tasks?. Applied Sciences, 15(6), 2971.
- Kim, S., Shin, J., Cho, Y., Jang, J., Longpre, S., Lee, H. & Seo, M. (2023, October). Prometheus: Inducing fine-grained evaluation capability in language models. In The Twelfth International Conference on Learning Representations.
- Kuhn, S. (2001). Learning from the architecture studio: Implications for project-based pedagogy. International Journal of Engineering Education, 17(4/5), 349-352.
- Kumar, H., Xiao, R., Lawson, B., Musabirov, I., Shi, J., Wang, X., Luo, H., Williams, J., Rafferty, A. N., Stamper, J., & Liut, M. (2024). Supporting Self-Reflection at Scale with Large Language Models: Insights from Randomized Field Experiments in Classrooms. Proceedings of the Eleventh ACM Conference on Learning @ Scale.
- Liu, B., Gui, W., Gao, T., Wu, Y., & Zuo, M. (2023). Understanding self-directed learning behaviors in a computer-aided 3D design context. Computers & Education, 205, 104882.

- Leicht, R. M., Messner, J. I., & Anumba, C. J. (2009). A framework for using interactive workspaces for effective collaboration. Journal of Information Technology in Construction (ITcon), 14(15), 180-203.
- Lodson, J., & Ogbeba, J. E. (2020). The Effect of Teacher-Student Relationships On Student Creative Performances in Architectural Design Studio. The Educational Review, USA, 4(2).
- Lotfabadi, P., & Iranmanesh, A. (2024). Evaluation of learning methods in architecture design studio via analytic hierarchy process: a case study. Architectural Engineering and Design Management, 20(1), 47-64.
- Louatouate, H., & Zeriouh, M. (2025). Meta-Prompting as a Solution to Students' Prompt Engineering Difficulties for an Optimized Use of GenAI LLMs in the Context of Education: A Quasi-Experimental Study using Mistral Model. Journal of Computer Science and Technology Studies, 7(2), 217-227.
- Lymer, G. (2009). Demonstrating professional vision: The work of critique in architectural education. Mind, Culture, and Activity, 16(2), 145-171.
- Mahmoud, N. E., Kamel, S. M., & Hamza, T. S. (2020). The relationship between tolerance of ambiguity and creativity in architectural design studio. Creativity Studies, 13(1), 179-198.
- Matzakos, N., & Moundridou, M. (2025). Exploring Large Language Models Integration in Higher Education: A Case Study in a Mathematics Laboratory for Civil Engineering Students. Computer Applications in Engineering Education, 33(3), e70049.
- McLaughlan, R., & Chatterjee, I. (2020). What works in the architecture studio? Five strategies for optimising student learning. International Journal of Art & Design Education, 39(3), 550-564.
- Megahed, N. (2017). Reflections on studio-based learning: assessment and critique. Journal of Engineering, Design and Technology.
- Mohanty, S. (2025). Fine-Grained Bias Detection in LLM: Enhancing detection mechanisms for nuanced biases. arXiv preprint arXiv:2503.06054.
- Murphy, K. M., Ivarsson, J., & Lymer, G. (2012). Embodied reasoning in architectural critique. Design Studies, 33(6), 530-556.
- Osmólska, D., & Lewis, A. (2023). Architects' use of intuition in site analysis: Information gathering in solution development.
- Özorhon, G., Nitelik Gelirli, D., Lekesiz, G., & Müezzinoğlu, C. (2025). AI-assisted architectural design studio (AI-a-ADS): How artificial intelligence join the architectural design studio?. *International Journal of Technology and Design Education*, 1-25.
- Pauwels, P., Strobbe, T., Derboven, J., & Meyer, R. (2014). Analysing the impact of constraints on decision-making by architectural designers.
- Park, E. J., Lee, K., & Kang, E. (2023). The impact of research and representation of site analysis for creative design approach in architectural design studio. Thinking Skills and Creativity, 48, 101271.
- Salama, A. (2016). Spatial design education: New directions for pedagogy in architecture and beyond. *Routledge.*

- Schon, D. A. (1987). Educating the Reflective Practitioner. Toward a New Design for Teaching and Learning in the Professions. The Jossey-Bass Higher Education Series. Jossey-Bass Publishers, 350 Sansome Street, San Francisco, CA 94104.
- Sinaga, Y. D. K., Arliani, E., Ngala, J. C., & Agustina, N. L. I. T. (2024). Accuracy of self-assessment and peer assessment in learning: A systematic literature review. Jurnal Paedagogy, 11(2), 312-322.
- Stanimirovic, M., Nikolic, B., Vasic, M., & Zivkovic, M. (2023). The role of visual thinking in educational development: architectural design. Journal of Asian Architecture and Building Engineering, 22(6), 3244-3252.
- Tschimmel, K. (2012). Design Thinking as an effective Toolkit for Innovation. In ISPIM conference proceedings (p. 1). The International Society for Professional Innovation Management (ISPIM).
- Wang, J. (2023). The comparison between architecture students' peer learning in informal situations within physical and virtual environments during the COVID-19 pandemic. Indoor and Built Environment, 32(10), 2064-2082.
- Wang, X., Li, C., Chang, Y., Wang, J., & Wu, Y. (2024). Negativeprompt: Leveraging psychology for large language models enhancement via negative emotional stimuli. arXiv preprint arXiv:2405.02814.
- Wang, J. (2025). Architecture students' peer learning in informal situations by lens of the community of practice – one case study. *Interactive Learning Environments*, *33*(7), 4394–4418. https://doi.org/10.1080/10494820.2025.2462152
- Xu, J., Wu, A., Filip, C., Patel, Z., Bernstein, S. R., Tanveer, R., & Kotroczo, T. (2024). Active recall strategies associated with academic achievement in young adults: A systematic review. Journal of Affective Disorders, 354, 191-198.
- Yaseen, A., Waqas, M., & Mukhtar, A. (2022). Precedent Study: An Approach to Learning about Design Challenges in Architectural Studio Pedagogy. Global Educational Studies Review, 7(1), 395-403.
- Yang, R., Ye, F., Li, J., Yuan, S., Zhang, Y., Tu, Z., ... & Yang, D. (2025). The lighthouse of language: Enhancing llm agents via critique-guided improvement. arXiv preprint arXiv:2503.16024.
- Yazici, S. (2020). Rule-based rationalization of form: learning by computational making. International Journal of Technology and Design Education, 30(3), 613-633.
- Yorgancıoğlu, D., & Tunali, S. (2020). Critique's Role in the Development of Design Literacy in Beginning Design Education. RChD: creación y pensamiento, 5(8), 49-62.
- Yue, M., Lyu, W., Mifdal, W., Suh, J., Zhang, Y., & Yao, Z. (2024). Mathvc: An llm-simulated multi-character virtual classroom for mathematics education. arXiv preprint arXiv:2404.06711.
- Yusoff, W. F. M., Ja'afar, N. H., & Mohammad, N. (2022). Preliminary investigation on architecture students' perceptions of developing hard and soft skills via project-based learning. Jurnal Kejuruteraan, 34(4), 629-637.

- Zhang, X., Tang, X., Liu, H., Wu, Z., He, Q., Lee, D., & Wang, S. (2024). Divide-verify-refine: Aligning llm responses with complex instructions.
- Zahra, F., Rana, M., & Rifat, A. (2025). Integration of artificial intelligence in architectural education: A SWOT-based study on students' perspectives. *Sustainability, 16*(5), 11135. https://doi.org/10.3390/su160511135
- Zahra, S., Samra, M., & El Gizawi, L. (2025). Working toward advanced architectural education: Developing an AI-based model to improve emotional intelligence in education. *Buildings, 15*, 356. https://doi.org/10.3390/buildings15030356
- Zhou, X., Li, R., Liang, P., Zhang, B., Shahin, M., Li, Z., & Yang, C. (2025). Using LLMs in Generating Design Rationale for Software Architecture Decisions. arXiv preprint arXiv:2504.20781.
- Zimmerman, B. (2002). Becoming a Self-Regulated Learner: An Overview. Theory Into Practice. *Taylor & Francis*, *41*(2), 64–70.